\documentclass[prb,twocolumn,showpacs]{revtex4-1}
\usepackage{graphicx}
\usepackage{amsmath}
\usepackage{amssymb}

\newcommand{\p}{\partial}
\newcommand{\pe}{\prime}

\newcommand{\eps}{\varepsilon}

\newcommand{\ra}{\rangle}
\newcommand{\la}{\langle}
\DeclareMathAlphabet{\bi}{OML}{cmm}{b}{it}
\begin{document}
%\title{Appearance of beating pattern in Weiss oscillation of spin-orbit 
%coupled Dirac fermions in gated silicene}
\title{Beating pattern in quantum magnetotransport coefficients of
spin-orbit coupled Dirac fermions in gated silicene}
\author{SK Firoz Islam and Tarun Kanti Ghosh}
\affiliation{Department of Physics, Indian Institute of Technology-Kanpur,
Kanpur-208 016, India}

\begin{abstract}
We report theoretical study of magnetotransport coefficients
of spin-orbit coupled gated silicene in presence and absence of
spatial periodic modulation. 
The combined effect of spin-orbit coupling and perpendicular electric 
field manifests through formation of regular beating 
pattern in Weiss and SdH oscillations.
Analytical results, in addition to the numerical results, of the beating
pattern formation are provided. 
The analytical results yield a beating condition which will be
useful to determine the spin-orbit coupling constant by simply counting  
the number of oscillation between any two successive nodes. 
Moreover, the numerical results of modulation effect on collisional and 
Hall conductivities are presented.

\end{abstract}
\date{\today}

\maketitle
\section{INTRODUCTION}

Silicene, a new class of two-dimensional (2D) electron system, 
possess graphene-like hexagonal lattice structure except the atoms 
are silicon instead of carbon\cite{takeda,guzman}. Several experiments 
of synthesizing monolayer silicene have been performed successfully 
\cite{lalmi,padova1,padova2,vogt,lin,PRL12}. 
Relatively larger atomic size of silicon atom causes the 2D lattice to be 
buckled, in which two planes of sublattice $A$ and sublattice $B$ are 
separated by $d \backsimeq 0.46 \mathring{A}$\cite{drummond,nano}.
Theoretical investigations\cite{drummond,liu1,liu2}
show that silicene possess an intrinsic spin-orbit coupling which is
very strong in comparison to graphene. 
This is due to the fact that {\mbox Si} atoms have large intrinsic spin-orbit 
coupling strength than {\mbox C} atoms. 
The application of electric field ($E_z$) perpendicular to the buckled 
silicene sheet generates a staggered sublattice potential difference 
($\Delta_z= E_z d$) between the two atomic planes and opens an electric field
dependent band gap between conduction and valence bands 
\cite{drummond,ni,ezawa}. 
The charge carriers in silicene also obey Dirac-like Hamiltonian\cite{vogt}
around the corners of its hexagonal Brillouin zone 
with additional properties of having intrinsic spin-orbit coupling and 
electrically tunable band gap\cite{drummond}.
Moreover, the low-energy dispersion around $K$ and $K^{\prime}$ points splits 
into two branches due to presence of the both spin-orbit coupling and $\Delta_z$.  
Monolayer graphene's zero band gap with the difficulty in tuning it and 
vanishingly small spin-orbit interaction prevent it to be used for electronic 
devices.
On the other hand, silicene overcomes all these limitations and 
provide an alternative to graphene for device based applications. 
Recently, a series of theoretical works on transport and optical properties of 
silicene have been reported, revealing the roles of finite band gap and 
spin-orbit interaction\cite{ezawa1,tahir_13,sabeeh_13,nicole,nicole1}.\\

The study of magnetotransport coefficients is one of the basic tools 
to investigate various two-dimensional electron systems. Appearance of 
quantum oscillation in magnetotransport coefficients, known as 
Shubnikov-de Hass(SdH) oscillation, 
is due to the interplay between Landau levels and Fermi energy.
A different kind of quantum oscillation,
known as Weiss oscillation, appears in magnetoresistance 
when an in-plane weak spatial electric modulation is applied 
\cite{Weiss,Gerh,kotha}.
The Weiss oscillation is due to the commensurability of the diameter of
the cyclotron orbit near the Fermi energy and the spatial period of the modulation
\cite{poulo,chao,vasilo,beenakker}.
Both the oscillations are periodic with inverse magnetic field.
The Weiss oscillation appears at very low magnetic field where
SdH oscillations are completely wiped out. On the other hand, at moderate
magnetic field, very weak Weiss oscillation is superposed on the SdH
oscillations.

The spin-orbit interaction lifts the spin degeneracy and 
produces two unequally spaced Landau levels for spin-up and 
spin-down electrons in a two-dimensional electron gas formed at
semiconductor heterostructures. The difference between two 
frequencies of quantum oscillation for spin-up and spin-down 
electrons is directly related to the spin-orbit coupling constant 
and yields beating pattern in the amplitude of the Weiss and SdH 
oscillations \cite{das,nitta,wang,firoz1,vasilo1,firoz2}. 
The beating pattern in the SdH oscillations is being 
used to determine the spin-orbit coupling constant. 
It is proposed 
that the beating pattern in Weiss oscillation can be also used to 
determine spin-orbit coupling constant \cite{firoz2}.

The SdH oscillation in graphene monolayer describing by massless 
Dirac-like Hamiltonian has been experimentally \cite{sdh_g,tahir}. 
The appearance of Weiss oscillation in graphene monolayer has been
predicted in Refs. \cite{matulis,tahir1}. 
The beating pattern in magnetotransport coefficients of 
graphene monolayer does not appear due to absent of spin-split Landau 
energy levels. On the other hand, spin-split Landau levels appear
in spin-orbit coupled Dirac fermions in gated silicene. 
There are several theoretical group estimated $\Delta_{\rm so} $ and
$\Delta_z$ using tight-binding and density functional calculations.
In this paper we discuss that $\Delta_{\rm so} $ and $\Delta_z$ can be
determined experimentally by analyzing beating pattern in magnetotransport 
coefficients in spin-orbit coupled gated silicene in presence and absence of
the spatial periodic modulation.

This paper is arranged in following order. In section II, we present  
energy eigenvalues, the corresponding eigenstates and density of states
of the charge carriers in silicene sheet subjected to 
transverse magnetic and electric fields. 
In section III, we present analytical and numerical results 
of diffusive, collisional and Hall conductivities in presence and
absence of the periodic modulation.  
In section IV we summarize our results. 

\section{ENERGY EIGENVALUE AND EIGENFUNCTION}
We are considering a buckled 2D silicene sheet in which Dirac 
electrons obey a finite gapped graphene-like Hamiltonian.
The Hamiltonian of an electron with charge $-e$ in presence of 
both field, the perpendicular magnetic field ${\bf B} $ and 
electric field ${\bf E}=E_z \hat{z}$, is\cite{liu2,ezawa1}
\begin{equation}
H = v_F(\sigma_x\Pi_x - \eta\sigma_y\Pi_y) - 
\eta s \Delta_{\rm so}\sigma_z + \Delta_z \sigma_z,
\end{equation}
where $v_{F}$ is the Fermi velocity, ${\bf \Pi} = {\bf p} + e{\bf A}$ 
is the 2D momentum operator with vector potential ${\bf A}$, 
$\eta= +(-)$ denotes $K(K^{\prime})$ Dirac point, $s=\pm$ stands for spin-up 
and spin-down, ${\mbox{\boldmath $\sigma$} }=(\sigma_x,\sigma_y,\sigma_z)$ 
are the Pauli matrices, $\Delta_{\rm so} $ is the strength of the 
spin-orbit interaction, and $\Delta_z$ is the energy associated with 
the applied electric field.

Using Landau gauge ${\bf A}=(0,Bx,0)$, the exact Landau levels 
and the corresponding wave functions are obtained in Refs. 
\cite{tahir_13,nicole}. The ground state energy ($n=0$) is 
$E_0^{\eta}=-(s\Delta_{\rm so}-\eta\Delta_z)$. 
For $n \geq 1$, energy spectrum for the electron band is
\begin{equation}
E_{\zeta} =\sqrt{n\eps^2+(\Delta_{\rm so}-\eta s\Delta_z)^2},
\end{equation}
where $\zeta \equiv \{n,s,\eta\}$, 
$\eps = \hbar\omega_c$ and $\omega_c= \sqrt{2}v_{F}/l$
is the cyclotron frequency with $l=\sqrt{\hbar/eB}$
is the magnetic length scale. 
The Landau levels around $K $ and $K^{\prime}$ points split into 
two branches due to presence of both $ \Delta_{\rm so} $ and $\Delta_z$.
The splitting vanishes when either $ \Delta_{\rm so} $ or $\Delta_z$
is zero.
The normalized eigenstates are (for $n \geq 1$)
\begin{equation}
\Psi^{\eta=+} _{n,s}({\bf r}) = 
\frac{e^{i k_y y}}{\sqrt{L_y N_{n,s}^{\eta}}}\left[
\begin{array}{r}
\alpha_{n,s}^{+} \phi_{n-1}(x + x_0)
\\
\beta_{n}\phi_n (x + x_0)
\end{array}
\right] \text{}
\end{equation}
and
\begin{equation}
\Psi^{\eta=-} _{n,s}({\bf r}) = 
\frac{e^{i k_y y}}{\sqrt{L_y N_{n,s}^{\eta}}}\left[
\begin{array}{r}
\alpha^{-}_{n,s} \phi_{n}(x + x_0)
\\
\beta_{n}\phi_{n-1} (x + x_0)
\end{array}
\right] \text{,}
\end{equation}
where  $\phi_n(x) = (1/\sqrt{\sqrt{\pi }2^n n!l}) 
e^{-x^2/2l^2} H_n(x/l) $ is the normalized harmonic 
oscillator wave function centred at $x=-x_0$ with 
$x_0=k_yl^2 $. Here, the coefficients are
$ N_{n,s}^{\eta}=\mid\alpha_{n,s}^{\eta}\mid^2+\mid\beta_n\mid^2 $
with
$\alpha^{\eta}_{n,s}=\Delta_{\eta,s}+\sqrt{\Delta_{\eta,s}^2 + 
n(\hbar\omega_c)^2} $
and
$ \beta_n=-i\hbar\omega_c\sqrt{n} $.
Here, $\Delta_{\eta, s}=(\Delta_{\rm so}-\eta s\Delta_z)$.

The analytical form of density of states \cite{dos} is given by
\begin{eqnarray} \label{dos_B}
D_{\eta,s}(E) & \backsimeq & D_{0}(E)
\Big[1+2\sum_{k=1}^{\infty} \exp\Big\{-k \Big(2\pi\frac{\Gamma_0 E }
{\eps^2}\Big)^2\Big\}
\nonumber\\
&\times&\cos\Big\{\pi s \Big(E^2-\Delta_{\eta,s}^2\Big)/\eps^2\Big\}\Big],
\end{eqnarray}
where $D_{0}(E)=E/(2\pi\hbar^2 v_{F}^2)$ 
and $\Gamma_0$ is the impurity induced Landau level broadening.

\section{Electrical Magnetotransport}
In this section we shall study magnetotransport coefficients
such as diffusive, collisional and Hall conductivities. The 
diffusive conductivity in absence of any modulation exactly 
vanishes because of the zero group velocity due to $k_y$ degeneracy 
in the energy spectrum. The presence of modulation imparts 
drift velocity to the charge carriers along the free direction of it's 
motion and gives rise to diffusive conductivity. The oscillations in the 
diffusive conductivity
known as Weiss oscillation which is dominant at low magnetic field regime.
On the other hand, the collisional conductivity arises due to scattering
of the charge carriers with localized charged impurities present in the 
system. The quantum oscillation in collisional conductivity is known as
SdH oscillation.
The Hall conductivity due to the Lorentz force is independent of 
any collisional mechanisms.
However, the external spatial periodic modulation induces periodic oscillation
on collisional as well as Hall conductivities.  
We shall use formalism of calculating 
different magnetotransport coefficients developed in Ref. \cite{vilet}.

\subsection{Diffusive conductivity}

To study diffusive conductivity, a spatial weak electrical 
modulation $V=V_0\cos(Kx)$ with $ K=2\pi/a$ is applied along 
the $x$ direction of the silicene sheet. Here, $a$ is the modulation 
period. We can treat this modulation as a weak perturbation as long as
$ V_0 \ll \epsilon $. The energy 
correction due to the modulation is calculated approximately by using 
first-order perturbation theory. Then total energy is given by
$ E_{\xi}= E_{\zeta}+\Delta E_{\xi}$,
where $ \xi \equiv \{\zeta,k_y\}$ and
$\Delta E_{\xi}= G_{\zeta}(u)\cos(Kx_0)$ with
\begin{equation}
G_{\zeta}(u)=\frac{V_0 e^{-u/2}}{N_{n,s}^{\eta}}
\Big[\mid\alpha_{n,s}^{\eta}\mid^2L_{n-1}(u) +
\mid\beta_n\mid^2L_n(u)\Big].
\end{equation}
Here, $u=(Kl)^2/2$ and $L_{n}(u)$ is the Laguerre
polynomial of degree $n$.
The energy correction $ \Delta E_{\xi} $  transforms the
degenerate Landau levels into bands due to $k_y$ dependency,
which leads to non-zero drift velocity.

The diffusive conductivity is calculated by using  the 
standard semiclassical expression as
\begin{equation} \label{conduc}
\sigma_{yy}^{\rm dif} = \frac{\beta e^2 \tau }{\Omega}
\sum_{\xi} f_{\xi}(1-f_{\xi}) (v_y^{\xi})^2,
\end{equation}
where $\Omega=L_x\times L_y$ is the area of the system, 
$\tau = \tau(E_{F})$ is the electron relaxation time at the 
Fermi energy $E_F$ which is calculated in the next paragraph, 
$ f_{\xi}$ is the Fermi-Dirac distribution 
function at $E=E_{\xi}$ and $\beta=(k_{B} T)^{-1}$ with $k_{B}$ 
is the Boltzmann constant. 
Also, $ v_y^{\xi} = \la \xi | \hat v_y | \xi \ra $ is 
the average value of the velocity operator $ \hat v_y $ and
it does not vanish due to the 
$k_y$ dependency of the energy levels. It is given by
\begin{equation} \label{vel}
v_y^{\xi} =  \frac{1}{\hbar} \frac{\partial E_{\xi}}{\partial k_y} = 
- \frac{K l^2}{\hbar} \sin(Kx_0) G_{\zeta}(u).
\end{equation}
After inserting drift velocity given by Eq. (\ref{vel}) into 
Eq. (\ref{conduc}) and integrating over $k_y $ variable, we get 
$ \sigma_{yy}^{\rm dif} = (e^2/h) \Phi $ with
\begin{equation}\label{con1}
\Phi=\frac{\beta \tau u}{\hbar} \sum_{\zeta}
f(E_{n,s}^{\eta})[1-f(E_{n,s}^{\eta})][G_{\zeta}(u)]^2
\end{equation}
is the dimensionless total diffusive conductivity. 
Note that a given Landau level splits into two branches due to the
simultaneous presence of $\Delta_{\rm so} $ and $\Delta_z$.
The first branch is 
$ E_n^{-} = \sqrt{n\eps^2+(\Delta_{\rm so} - \Delta_z)^2}$ 
for $\{s,\eta\}=\{+,+\}$ and $\{-,-\}$. 
The second branch  is 
$ E_n^{+} = \sqrt{n\eps^2+(\Delta_{\rm so} + \Delta_z)^2}$ 
for $\{s,\eta\}=\{+,-\}$ and $\{-,+\}$.
So, for $\eta=+1$ ($K$-valley) there are two energy branches 
due to spin splitting and same goes for $\eta=-1$ ($K^{\prime}$-valley) 
also. 
The total conductivity is written as $ \Phi=\Phi^{+} + \Phi^{-}$.
Here, $\Phi^{+}$ is the contribution from the second branch $E_n^+$ 
and $\Phi^{-}$ is coming from the first branch $E_n^-$.

\begin{figure}[t]
\begin{center}\leavevmode
\includegraphics[width=98mm]{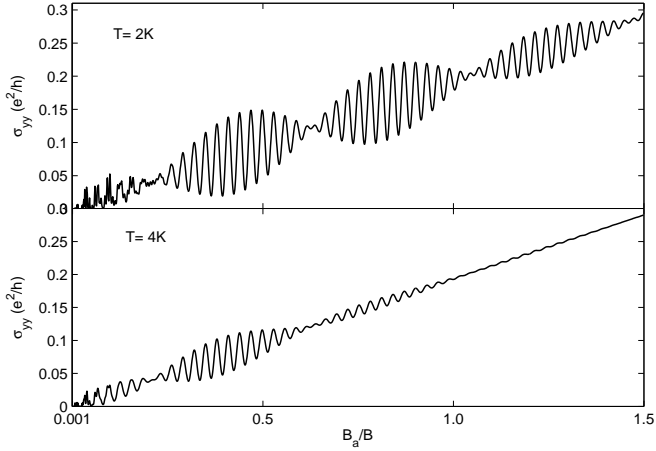}
\caption{Plots of the exact results of the diffusive
conductivity versus dimensionless inverse magnetic field $\lambda$.}
\label{Fig1}
\end{center}
\end{figure}

Before simplifying further, we shall derive the Fermi energy.
At Fermi level $ E_{F}=\sqrt{(\hbar v_F k_F^{\pm})^2 + 
\Delta^{2}_{\pm}}$ with $\Delta_{\pm}=(\Delta_{\rm so}\mp\Delta_z)$, 
then one can write
\begin{equation}
(k_F^{+})^2 - (k_F^{-})^2 = 
4\frac{\Delta_{\rm so}\Delta_z}{(\hbar v_F)^2}.
\end{equation}
On the other hand, carrier density is given by
\begin{equation}
n_e=\frac{2}{(2\pi)^2} \int_0^{2\pi}\int_{0}^{k_F}k dk d\phi = 
\frac{1}{2\pi} \Big[(k_{F}^{+})^2 + (k_{F}^{-})^2 \Big].
\end{equation}
Solving the above two equations for Fermi energy, we get 
$E_F=\sqrt{(E_{F}^0)^2+\Delta_{\rm so}^2 + \Delta_z^2}$
with $E_{F}^0=\hbar v_{F} k_{F}^0$ and $k_{F}^0=\sqrt{\pi n_e}$.

Before presenting exact analytical results, we would like to 
get approximated analytical results. To do so we consider the system  
at very low temperature in which Landau levels close to the Fermi energy
contribute to transport properties. Therefore, we can use some 
approximations which are valid for higher values of $n$. 
Around Fermi level, for higher value of $n$, we have
\begin{equation}
e^{-u/2} L_n(u) \simeq \frac{1}{\sqrt{\pi \sqrt{nu}}} 
\cos \Big(2\sqrt{nu} - \pi/4 \Big). \label{limit}
\end{equation}
We can also use $n\backsimeq n-1$ for higher values of Landau levels, 
which give
$ G_{\zeta}(u)\simeq V_0\exp (-u/2 ) L_{n}(u)$.
To obtain analytical expression we convert the summation into integration
by using the relation 
\begin{equation}
\sum_{n}\rightarrow\int_0^{\infty} dn 
\simeq\frac{2}{\eps^2}\int_0^{\infty}EdE.
\end{equation}

By using the above two approximations, the exact expression 
of the dimensionless diffusive conductivity given by Eq. (\ref{con1}) 
reduces to  
\begin{eqnarray}\label{con2}
\Phi^{\pm} & \backsimeq & A_0\frac{u}{c^{\pm}}
\int_{-\infty}^{\infty}\frac{\cos^2(c^{\pm}t+d^{\pm})}
{\cosh^2(t/2)}dt,\nonumber\\
\end{eqnarray}
where $A_0=V_0^2\tau_0/(\hbar\beta\eps^2)$, $t=\beta(E-E_{F})$, and 
$ d^{\pm}=c^{\pm}\beta E_F $ with
\begin{equation}
c^{\pm}=\frac{2\sqrt{u}}{\eps}\frac{\sqrt{(E_{F}^0)^2\mp2\Delta_{\rm so} 
\Delta_z}}{\beta E_F}.
\end{equation}                                                        

Using the standard integral \cite{inti}, Eq. (\ref{con2}) reduces
to
\begin{equation}
\Phi^{\pm}=A_0\frac{u}{c^{\pm}}
\Big[1+H\Big(T/T_a^{\pm} \Big)\cos\Big(2\pi f^{\pm}\lambda\Big)\Big].
\end{equation}
Here, $ \lambda=B_a/B$ with $B_a=\hbar/ea^2$ and 
\begin{equation}
f^{\pm}=\frac{2a}{\hbar v_F}\sqrt{(E_{F}^0)^2\mp2\Delta_{\rm so}\Delta_z}
\end{equation}
are two closely spaced frequencies of Weiss oscillations of the two 
energy branches induced by the simultaneous presence of spin-orbit
interaction and gate induced electric field.
The temperature dependent damping factor is given by 
\begin{eqnarray}
H\Big(T/T_a^{\pm}\Big) & = 
&\frac{T/T_a^{\pm}}{\sinh(T/T_a^{\pm})}\nonumber,
\end{eqnarray}
where the characteristic temperature $(T_a^{\pm})$ is 
defined as
\begin{equation}\label{chrt}
T_a^{\pm}=\hbar v_F E_{F}B/\Big[4\pi^2k_{B}aB_{a} 
\sqrt{(E_{F}^0)^2\mp2\Delta_{\rm so}\Delta_z}\Big].
\end{equation}
Typically the difference between $T_a^{+}$ and $T_a^{-}$ is very small. 
$T_a^{-}=0.76$ K when $B=0.6$ T. 
Another point is that $T_a^{+}$ is increasing with  electric field ($\Delta_z$)
or spin-orbit interaction ($\Delta_{\rm so}$) while $T_a^{-}$ is decreasing.
\begin{figure}[t]
\begin{center}\leavevmode
\includegraphics[width=98mm]{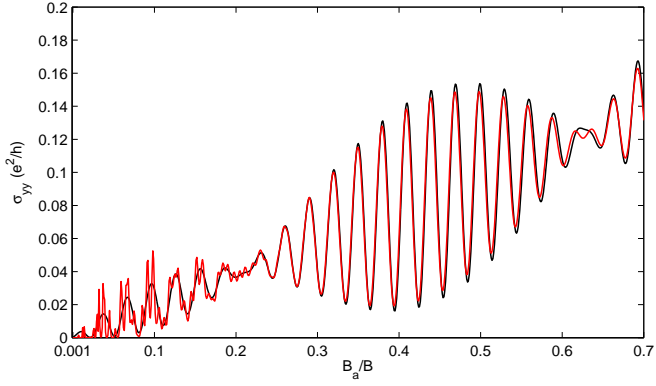}
\caption{Plots of the exact and analytical results of diffusive
conductivity versus dimensionless inverse magnetic field. 
The dark and red lines stand for analytic and numerical results.}
\label{Fig1}
\end{center}
\end{figure}

As $(E_{F}^0)^2 \gg 2\Delta_{\rm so} \Delta_{z}$, 
the total diffusive conductivity is given by
\begin{equation} \label{beat}
\Phi \backsimeq 2A_{0}\frac{u}{c}\big[1+H\big(T/T_a\big) 
\cos(2\pi f_{\rm av}\lambda)\cos(2\pi f_{d}\lambda)\big]
\end{equation}
with $c=2\sqrt{u}E_{F0}/(\beta E_{F})$, 
$T_a=\hbar v_F E_{F}B/(4\pi^2k_{B}aB_{a}E_{F}^0)$,
$f_{\rm av}=(f^{+}+f^{-})/2 $ and  $f_{d}=(f^{+}-f^{-})/2 $.
The total diffusive conductivity given by Eq. (\ref{beat}) exhibits 
beating pattern due to the superposition of two oscillatory functions
with closely spaced frequencies $f^{\pm}$. It should be noted that 
the condition $(E_{F}^0)^2 > 2\Delta_{\rm so}\Delta_z$
must be satisfied to see the beating pattern.
The location of the beating node can be obtained from the condition
$\cos(2\pi f_{d}\lambda)\mid_{_{B=B_j}}=0$, which 
gives 
\begin{eqnarray} \label{node}
2f_d\frac{B_a}{B_j}=\Big(j+\frac{1}{2}\Big),
\end{eqnarray}
where $ j = 0,1,2...$.
Another periodic term, $\cos(2\pi f_{\rm av}\lambda)$, gives 
number of oscillation between two successive beating nodes as
\begin{equation}\label{beat2}
N_{\rm osc}=f_{\rm av}\Big(\frac{B_a}{B_{j+1}} - 
\frac{B_a}{B_j}\Big) = \frac{1}{2}\frac{f_{\rm av}}{f_d}.
\end{equation}
In explicit form, it is given by
\begin{equation}\label{beat_eq}
2N_{\rm osc} = \frac{\sqrt{(E_{F}^0)^2 + 
2\Delta_{\rm so}\Delta_z} + \sqrt{(E_{F}^0)^2 - 
2 \Delta_{\rm so}\Delta_z}}
{\sqrt{(E_{F}^0)^2 + 2\Delta_{\rm so}\Delta_z} - 
\sqrt{(E_{F}^0)^2 - 2\Delta_{\rm so}\Delta_z}}.
\end{equation}
Here we make couple of important remarks on the above equation:
1) number of oscillation between any two successive beat nodes is 
independent of the modulation period $a$ and magnetic field 
and 2) the expression of $N_{\rm osc}$ can be used to estimate 
the spin-orbit coupling constant by simply counting the number of 
oscillations between any two successive beat nodes.

We use the following parameters for various plots: 
electron density $n_e = 4 \times 10^{15}$m${}^{-2}$, spin-orbit coupling 
constant $\Delta_{\rm so}=4$ meV, perpendicular electric field
induced energy $\Delta_z=12 $ meV, 
Fermi velocity $v_{F}=2\times 10^5$ ms${}^{-1}$, 
modulation period $a=150$ nm, and modulation strength $V_0=0.1$ meV.
Figure 1 shows exact numerical results of the diffusive conductivity given
by Eqs. (\ref{conduc}) and (\ref{vel})  for two different temperature. 
It shows beating pattern in both the cases, and
oscillation gets damped with increasing temperature. 
In Fig. 2 we compare the approximate analytical result given by Eq. (\ref{beat}) 
with the exact numerical result obtained from Eq. (\ref{conduc}). 
The analytical result for diffusive conductivity matches very well with 
the exact result. 
The appearance of beating pattern is due to the superposition
of Weiss oscillation coming from two oscillatory drift velocity
with different frequencies ($f^{\pm}$) corresponding to two energy branches.
For the numerical parameters used here, the number of oscillations 
calculated from Eq. (\ref{beat_eq}) is 14, same as counted from 
Fig. 1.

\subsection{Collisional Conductivity}
First we shall study collisional conductivity in absence of
modulation. The effect of modulation on SdH oscillation will
be discussed in the later half of this section.
The standard expression for collisional conductivity is given by\cite{vilet}
\begin{equation} \label{coll}
\sigma_{\mu \mu}^{\rm col} = \frac{\beta e^2}{2\Omega}
\sum_{\zeta, \zeta^{\prime}} f_{\zeta} (1-f_{\zeta^{\prime}}) 
W_{\zeta, \zeta^{\prime}} 
(\alpha_\mu^{\zeta} - \alpha_{\mu}^{\zeta^{\prime}})^2.
\end{equation}
Here, $f_{\zeta}=f_{\zeta^{\prime}}$ for elastic scattering, 
$ W_{\zeta,\zeta^{\prime}} $ is the transition probability between 
one-electron states $ |\zeta \ra $ and $ | \zeta^{\prime} \ra $. 
Also, $ \alpha_{\mu}^{\zeta} = \la \zeta | r_{\mu} | \zeta \ra $ is
the average value of $\mu$ component of the position 
operator of the charge carriers in state $ |\zeta \ra $.
The scattering rate $ W_{\zeta,\zeta^{\prime}} $ is given by
\begin{equation}
W_{\zeta,\zeta^{\prime}} = \sum_{{\bf q}} | U ({\bf q})|^2 |\la \zeta | 
e^{i {\bf q} \cdot ({\bf r } - {\bf R})}| \zeta^{\prime} \ra|^2 
\delta(E_{\zeta} - E_{\zeta^{\prime}}),    
\end{equation}
where $ {\bf q}=q_{_{x}} \hat x + q_{_{y}} \hat y$ is 
the two-dimensional wave-vector and 
$ U({\bf q}) = 2 \pi e^2/
(\epsilon \sqrt{q_{_{x}}^2 + q_{_{y}}^2 + k_s^2})$ 
is the Fourier transform of the screened impurity potential 
$ U({\bf r}) = (e^2/4\pi \epsilon) (e^{-k_sr}/r) $, where
$k_s$ is the inverse screening length and $ \epsilon $ is 
the dielectric constant of the material. 
Equation (\ref{coll}) can be re-written for higher values 
of Landau level ($n \backsimeq n-1$) as given by
\begin{equation}
\sigma_{xx}^{\rm col}=\frac{e^2}{h}\frac{N_IU_0^2}{\pi l^2\Gamma_0}
\sum_{\zeta}(2n+1)\Big[-\frac{\partial f}{\partial E}
\Big]_{E=E_{\zeta}}.
\end{equation}
A closed-form analytical expression of the above equation is 
obtained by replacing the summation over quantum number as 
$\sum_n \rightarrow 2\pi l^2\int_0^\infty D(E) dE $
with $D(E)$ is the density of states (see Eq. (\ref{dos_B})) and it is given by
\begin{eqnarray}
\sigma_{xx}^{\rm col(\pm)} & \backsimeq & 
\frac{\sigma_0}{(\omega_c\tau_0)^2}
\frac{E_F^2-\Delta_{\pm}^2}{\eps^2}\nonumber\\ 
& \times & \Big[1+2\Omega_DH\Big( T/T_c \Big)
\cos\Big(2\pi \nu^{\pm}/B \Big)\Big].
\end{eqnarray}
Here, $ \sigma_{0}=e^2\tau_{0}E_{F}/\pi\hbar^2$ is Drude-like
conductivity and
$\nu^{\pm} = [(E_{F}^0)^2\pm 2\Delta_z\Delta_{\rm so}]/(2e\hbar v_F^2) $
are the SdH oscillation frequencies for the two energy branches $E_n^+ $ 
and $E_n^-$.
Also, impurity induced damping factor is
\begin{equation}
\Omega_{D}=\exp\Big\{-\Big(2\pi\frac{\Gamma_0 E_F }
{\eps^2}\Big)^2\Big\}
\end{equation}
and the temperature dependent damping factor is
$ H(x)=x/\sinh(x)$. Here, $x=T/T_c$ with $T_c=\eps^2/(2\pi^2k_{B}E_{F})$
is the critical temperature.

Following the same procedure as in the diffusive conductivity case, here
we get the location of a beating node ($B_j$) as 
$ B_j (j+1/2) = 2 \Delta_{\rm so} \Delta_z/(e \hbar v_F^2) $ and 
number of oscillations between any two successive nodes as 
$ N_{\rm osc}^{\rm s}= (E_{F}^0)^2/(4 \Delta_{\rm so}\Delta_z)$.
Unlike Rashba spin-orbit coupled two-dimensional electron gas, 
$ N_{\rm osc}^s $ in silicene is same for a set of given parameters and 
does not depend on specific choices of the successive nodes.
In Fig. 3, we show the beating pattern in the collisional conductivity 
vs inverse magnetic field. For the parameters used here, we have 
$N_{\rm osc}^s = 14$ which is same as shown in Fig. 3.

\begin{figure}[t]
\begin{center}\leavevmode
\includegraphics[width=90mm,height=44mm]{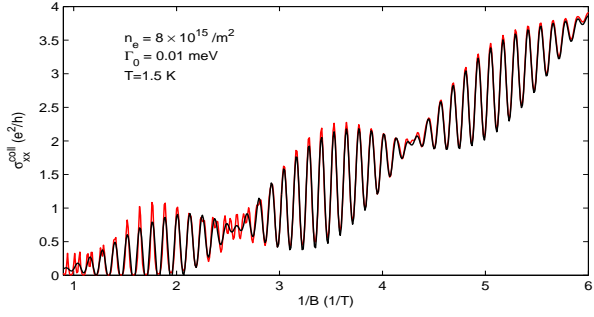}
\caption{Plots of the exact and analytical results of the collisional
conductivity versus inverse magnetic field in absence of modulation.
Here, red and dark lines stand for numeric and analytic results, respectively.}
\label{Fig1}
\end{center}
\end{figure}

Now we will describe effect of of weak modulation on the collisional 
conductivity. The modulation effect enters mainly through 
the total energy in Fermi distribution function.
The collisional conductivity in presence of weak modulation is given
by
\begin{equation}\label{collisional}
\sigma_{xx}^{\rm col} \backsimeq \frac{e^2}{h}\frac{N_IU_0^2}{\pi a\Gamma_0}
\sum_{\zeta}(2n+1) M_{\zeta},
\end{equation}
where $M_\zeta$ is given by
\begin{eqnarray}
M_{\zeta}&=&\int_{0}^{a/l^2}\Big[-\frac{\partial f}{\partial E}
\Big]_{E=E_{\xi}}dk_y.
\end{eqnarray}
It is difficult to get a closed-form analytical expression of the collisional 
conductivity in presence of the modulation. 
The change in collisional conductivity due to the modulation
[$ \Delta \sigma_{xx}^{\rm coll} = \sigma_{xx}^{\rm coll}(V_0)
- \sigma_{xx}^{\rm coll}(0)]$
is calculated from Eq. (\ref{collisional}) numerically and  shown in Fig. 4. 
\begin{figure}[t]
\begin{center}\leavevmode
\includegraphics[height=46mm, width=92mm]{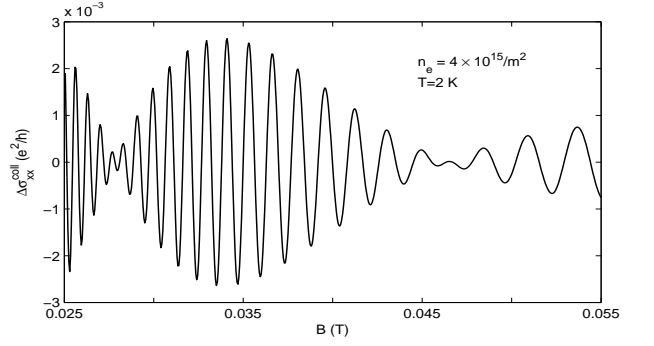}
\caption{Plots of change in the collisional conductivity 
due to modulation versus magnetic field.}
\label{Fig1}
\end{center}
\end{figure}
Figures (3) and (4) clearly shows that effect of the modulation
on $ \sigma_{xx}^{\rm coll} $ is very small and vanishes with increasing $B$.
The location of the beating node and number of oscillations between
any two successive nodes are determined by Eqs. (\ref{node}) 
and (\ref{beat2}), respectively.
To understand why the beating pattern appears in $ \Delta \sigma_{xx}^{\rm coll}$  
follows the same condition as in the diffusive conductivity,
we expand $M_{\zeta}$ as given by
\begin{eqnarray}
M_{\zeta}&=&\int_0^{a/l^2}\Big[\Big(-\frac{\p f}{\p E}\Big)+
\Delta E_{\zeta,k_y}\Big(-\frac{\p f^{\pe}}{\p E}\Big)\nonumber\\ 
&+& \frac{(\Delta E_{\zeta,k_y})^2}{2!} 
\Big(-\frac{\p f^{\pe\pe}}{\p E}\Big) + 
.....\Big]_{E=E_{\zeta,k_y}}dk_y \nonumber \\ 
& = &\frac{a}{l^2}\Big[\Big(-\frac{\p f}{\p E}\Big) +
\frac{\{G_{\zeta}(u)\}^2}{4}\Big(-\frac{\p f^{\pe\pe}}{\p E}\Big)...
\Big]_{E=E_{\zeta}}.
\end{eqnarray}
We can see the modulation dependent dominant term is of the order 
of $ (G_{\zeta}(u))^2 \simeq  V_0^2$, which is same as in the
diffusive conductivity (see Eq. \ref{con1}). The modulation induced Weiss 
oscillation in collisional conductivity, as shown in Fig. 4., also 
follows the same beating condition as in the diffusive conductivity.

From the numerical result we can see that modulation effect is dominant at low
range of magnetic field i.e; when the energy scale of Landau level is not much
higher than the energy correction due to modulation. 
As magnetic field increases, SdH oscillation starts to dominate over 
Weiss oscillation.

\subsection{The Hall conductivity} 
In this sub section, we will see modulation effect on the Hall
conductivity. The Hall conductivity is given by \cite{vilet}
\begin{eqnarray}\label{hall} 
\sigma_{yx} &=& \frac{i e^2 \hbar}{\Omega} 
\sum_\xi  \frac{f_{\xi}-f_{\xi^{\prime}}}
{{(E_{\xi} - E_{\xi^{\prime}})^2}}
\la \xi| \hat{v}_y|\xi^{\prime}\ra \la \xi^{\prime}| \hat{v}_x | \xi \ra.
\end{eqnarray}

\begin{figure}[t]
\begin{center}\leavevmode
\includegraphics[width=90mm]{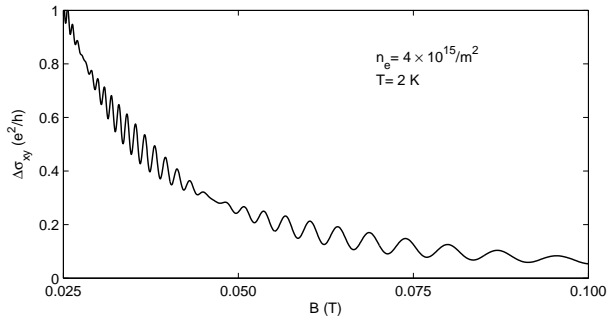}
\caption{Plots of the modulation effect on Hall
 conductivity versus magnetic field.}
\label{Fig1}
\end{center}
\end{figure}
Using unperturbed eigenstates velocity matrix elements are given by
\begin{equation} \label{vme}
<n,s,+\mid\hat{v}_{x}\mid n+1,s,+> = 
\frac{\alpha_{n+1,s}^{+}\beta_{n}}{N_{n+1,s}^{+} N_{n,s}^{+}}v_{F}
\end{equation}
and
\begin{equation} \label{vme1}
<n+1,s,+\mid\hat{v}_{y}\mid n,s,+> = 
-i\frac{\alpha_{n+1,s}^{+}\beta_{n}}{N_{n+1,s}^{+} N_{n,s}^{+}}v_{F}.
\end{equation}
Substituting Eqs. (\ref{vme}) and (\ref{vme1}) into Eq. (\ref{hall}),
we get
\begin{eqnarray}
\sigma_{yx}&=&2\frac{e^2}{h}\frac{l^2}
{a}\sum_{n,s}\int_0^{a/{l}^2}\frac{\mid\alpha_{n+1,s}^{+} 
\mid^2\mid\beta_{n}\mid^2}{\mid N_{n+1,s}^{+}\mid
\mid N_{n,s}^{+}\mid}
\nonumber\\&\times&\frac{f(E_{n,s,k_y}^{+})-f(E_{n+1,s,k_y}^{+})}
{[\sqrt{n+z}-\sqrt{n+1+z} -\rho_{n,k_y}]^2}dk_y.
\end{eqnarray}
Here $z=[(\Delta_{\rm so}-s\Delta_z)/\eps]^2$ and 
\begin{eqnarray}
\rho_{n,k_y} & = & \Delta E_{n+1,s,k_y}^{\eta}-\Delta E_{n,s,k_y}^{\eta} 
\nonumber \\ 
& \simeq & \frac{V_0}{\eps}\frac{u}{n}e^{-u/2}L^{1}_{n-1}\cos{(Kx_0)}.
\end{eqnarray}

Here, a factor of 2 is multiplied because of the 
identical states between two valleys but with opposite spin. 
The modulation induced change in the Hall conductivity 
$ [\Delta \sigma_{yx} = \sigma_{yx}(V_0) - \sigma_{yx}(0)] $ 
is plotted in Fig. 5. It shows beating pattern in the Weiss 
oscillation of the Hall conductivity. 
The node position $B_j$ and number of oscillations between
any two successive nodes are determined by Eqs. (\ref{node})
and (\ref{beat2}), respectively. 
The increase of magnetic field
diminishes the modulation effect on Hall conductivity as 
expected.

\section{SUMMARY}

We have shown the appearance of beating pattern in quantum 
oscillations of magnetotransport coefficients of spin-orbit 
coupled gated silicene with and without spatial periodic 
modulation. There is a spin-splitting of the Landau energy 
levels due to the presence of both the spin-orbit coupling 
and electric field perpendicular to the silicene sheet. The 
formation of beating pattern is due to the superposition of 
oscillations from two different energy branches but with slightly
different frequencies depending on the strength of spin-orbit
coupling constant and perpendicular electric field.
In addition to the numerical results we also provide 
analytical results of the beating pattern in Weiss and SdH oscillations.
The approximated analytical results are in excellent agreement with
the exact numerical results.
The analytical results yields a simple equation 
which can be used to determine the strength of spin-orbit coupling 
constant by simply counting the number of oscillations between any 
two successive beat nodes.
There have been few theoretical calculation, already mentioned 
earlier, which estimated the spin-orbit coupling constant. 
Here we have proposed a way to determine 
the spin-orbit coupling constant experimentally.
Finally for the sake of completeness, modulation effect on collisional
and Hall conductivities are also studied numerically.
Moreover, the analytical results of the Weiss and SdH oscillations 
frequencies reduce to  graphene monolayer case \cite{matulis,tahir1} 
by setting $\Delta_{\rm so} =0 $ or $\Delta_z = 0 $.

\section{acknowledgement}
This work is financially supported by the CSIR, Govt. of India under 
the grant CSIR-JRF-09/092(0687) 2009/EMR F-O746.

\end {document}